\documentclass{article}

\usepackage{arxiv}

\usepackage[utf8]{inputenc} % allow utf-8 input
\usepackage[T1]{fontenc}    % use 8-bit T1 fonts
\usepackage{hyperref}       % hyperlinks
\usepackage{url}            % simple URL typesetting
\usepackage{booktabs}       % professional-quality tables
\usepackage{amsfonts}       % blackboard math symbols
\usepackage{nicefrac}       % compact symbols for 1/2, etc.
\usepackage{microtype}      % microtypography
\usepackage{lipsum}		% Can be removed after putting your text content
\usepackage{graphicx}
\usepackage{subfig}
\usepackage{amsmath} 
\usepackage{epsfig}
\usepackage{breqn}
\usepackage{epstopdf}
\usepackage{bm}
\usepackage{amssymb}
\usepackage{color}
\usepackage{multirow}

\title{Impact of Inference Accelerators on Hardware Selection}

\author{ Dibyajyoti Pati,~ Caroline Favart,~ Purujit Bahl,~ Vivek Soni,~ Yun-chan Tsai,~ \And
Michael Potter,~ Jiahui Guan,~ Xiaomeng Dong,~ V. Ratna Saripalli\\ \\GE Healthcare }

\begin{document}

\maketitle

\begin{abstract}
As opportunities for AI-assisted healthcare grow steadily, model deployment faces challenges due to the specific characteristics of the industry. The configuration choice for a production device can impact model performance while influencing operational costs. Moreover, in healthcare some situations might require fast, but not real time, inference. We study different configurations and conduct a cost-performance analysis to determine the optimized hardware for the deployment of a model subject to healthcare domain constraints. We observe that a naive performance comparison may not lead to an optimal configuration selection. In fact, given realistic domain constraints, CPU execution might be preferable to GPU accelerators. Hence, defining beforehand precise expectations for model deployment is crucial.
\end{abstract}

\section{Introduction}
Over the last 10 years, healthcare demand for AI has exploded and deep learning technologies are being used to build cutting-edge algorithms to assist health professionals with their work. Once trained and evaluated, these models need to be deployed on the right devices to meet performance requirements, both in terms of precision and inference speed.

Healthcare AI applications are very specific, and models are often deployed within the context of long-running workflows. For instance, a Magnetic Resonance Imaging (MRI) procedure lasts 45 minutes to 1 hour on average per body part, so instantaneous inferencing is not required in order to achieve real-time results. Time remains a precious resource and optimizing inference speed is important, but its utility may be bounded by end-user context. Such limits become decision variables for hardware selection.

Model deployment should focus on choosing the most optimized hardware for a specific use-case. Identifying this optimal configuration based on the task to solve, the data that will be used, and other constraints such as inference speed, is a difficult challenge. For example, speed can vary dramatically between a CPU and GPU but specialized hardware is pricy and cost-to-performance ratio will be an important metric for any hardware selection process in a commercial context.

Our goal is to identify the optimal deployment environment given a specific model architecture, input data, and time constraints. We study different hardware performances for inferencing and conduct a price-to-performance analysis. We also analyze situations with specific hardware constraints and other challenges impacting deployment decisions. We focus on inference and not training, where the resource requirements are different. The analysis is performed for image processing within the healthcare domain, which to the best of our knowledge is the first such work to span multiple devices and model architectures.

\section{Methods}
In our work, we analyze cost-to-performance trade-offs to inform the selection of an optimal hardware platform and inferencing accelerator thereon. We perform our experiments on three different hardware configurations: NVIDIA K80 GPU, NVIDIA V100 GPU, and Intel Xeon CPU (see Table 1 in Appendix for configuration details). Furthermore, we extend our experiments to both Linux and Windows Operating Systems (OS) to characterize any variation arising from the choice of OS. In order to make the comparisons fair, we ensure that the inference time excludes disk I/O operations. For inferencing accelerators we focus on OpenVINO \cite{OpenVINO} (CPU), TensorRT \cite{TRT} (GPU), and WinML \cite{wml} (CPU, GPU).

Our method involves running a predefined set of experiments on competing configurations and selecting the configuration that minimizes the total inference cost thereon. The total inference cost is the product of the inference time for a million images and the cost of the hardware per unit time, subject to the constraint of completing the inference within a time bound defined by the user.  In some situations additional constraints on the model footprint can be relevant; however, in this analysis we focus on inference time. Even though hardware decisions cover several other costs, especially when considering on premise versus cloud, we limit our comparison to runtime costs only.

In order to establish the cost of hardware per unit time, we analyze the price offered by several cloud providers for the hardware configuration under analysis (or closely similar configuration) and establish an average. We tried to ensure that all other parameters of the systems such as memory and IO are comparable. Please refer to Table \ref{inference-table1} for the prices used in our analysis.

We conducted experiments with two models: InceptionV3 \cite{szegedy2016rethinking} with input shape $299 \times299 \times3$ for classification, and a Unet \cite{ronneberger2015u} model trained for healthcare image segmentation, with an input data shape of $160 \times160 \times1$. The dataset was comprised of 1000 images for the classification task and 40 images for the segmentation task. The dataset was handpicked to maintain a uniform distribution across different classes and to be representative of real-life difficulty in prediction. Furthermore, in order to eliminate system-related variations during experimentation, we randomize the inputs and run each trial for several thousand iterations. These experiments help us observe potential differences in results due to variations in hardware platforms, tasks, etc.

It is worth noting that inferencing accelerators do not support many commonly used neural network layers and therefore may require designing custom layers when converting models. For instance, “ResizeNearestNeighbor” and “ResizeBilinear” interpolations are not supported by TensorRT and OpenVINO. ResizeNearestNeighbor and ResizeBilinear are widely adopted interpolation strategies for up-sampling an image and are quite commonly used in segmentation, de-noising and super-resolution models.

\section{Results}
All models across devices maintain the same output performance\textemdash accuracy for classification and dice score for segmentation\textemdash except for Floating Point 16 (FP16) inferencing on V100 where the accuracy drops by 1\% for InceptionV3. The inference time was reduced by an average of 55\% from TensorFlow baseline for both classification and segmentation tasks, without any hardware accelerators (refer to Figure 1 in Appendix). For some medical domain tasks, such an improvement may be sufficient. When investigating inference accelerators; however, we observed an undesirable behavior originating from model conversion. When using the latest version of TensorRT (version 5.1.5, CUDA 10.0 and CuDNN 7.5), we saw a drop in dice score for our segmentation model. Downgrading it to 5.1.2 RC resolved this issue. It illustrates the need for caution while accelerating and converting models to avoid losses in absolute performance. To check the model output differences, we conducted a statistical hypothesis test with $H_{o}$: The true mean difference between the paired model outputs is zero. Given that the sample size is relatively small and the distribution is not normal, we employed a paired Wilcoxon Rank Sum test \cite{wilcoxon}. The p-values are all over 0.05 for TensorRT and OpenVino. We therefore accept the null hypothesis that the model outputs are statistically equivalent between OpenVINO and TensorRT across different machines. 

Both models have higher inference speeds on an NVIDIA V100 GPU than either of the other configurations. Although the GPU always provides faster inference, the difference in total inference costs (\$/image) between CPU inference and inferring on an NVIDIA K80 is not as high as the one between NVIDIA K80 and NVIDIA V100 (FP32) configurations. Indeed, inferring with the CPU is only 10\% slower on Intel Xeon compared to NVIDIA K80 for InceptionV3, whereas K80 is 74\% slower than V100. This difference is significant enough to cover the price differences between the two GPUs, making the V100 a more suitable option. This also points to the fact that GPU architecture (generation of GPU) plays a more vital role than GPU memory when compared from the perspective of cost-to-performance ratio. 

This gap between the older K80 and newer V100 further increases when we infer using half precision (FP16) on the V100, a feature which is not available on K80. It is quite intriguing to note that using FP16 for inferencing drops the overall inference cost on the V100 by an average of 38\% across both classification and segmentation tasks when compared with a K80 as shown in Table \ref{inference-table1}. This also reduces the gap between V100 and the cheaper Xeon configuration. It should be noted that FP16 mode may lead to a drop in performance in certain cases. During our experimentation we noted that for our classification task, FP16 mode resulted in a drop of 1\% in accuracy. Although this might seem marginal, it can be a vital factor for sensitive healthcare applications. With the FP16 support available on newer GPU cards, it becomes a pivotal consideration in decision making as it can bring down the total inferencing cost.

\begin{figure}[t]
  \centering
  \subfloat[]{\includegraphics[trim=0cm 0cm 0cm 0cm, clip, width=2.5in]{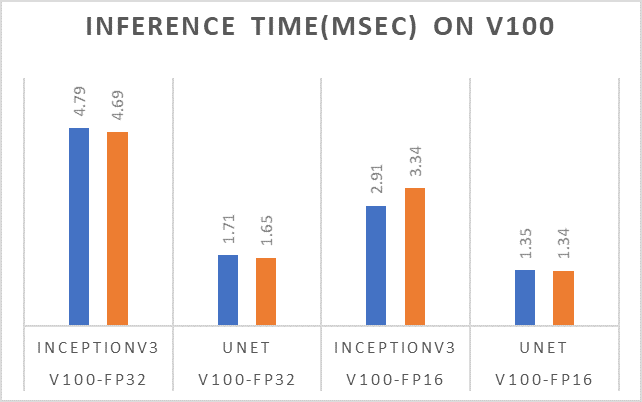}} 
  \subfloat[]{\includegraphics[trim=0cm 0cm 0cm 0cm, clip, width=2.5in]{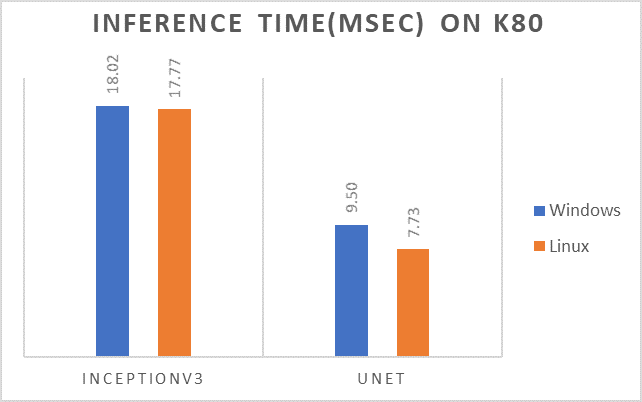}} 
  \caption{{Inference time comparison on (a) V100 and (b) K80. }}
  \label{fig:f2}
\end{figure}

We performed experiments on Linux and Windows OS. While accuracy was consistent across both OS, inference speed is higher on Linux: the inference time increased on average by 2\% for InceptionV3 and 15\% for Unet for experiments conducted on a GPU as shown in Figure \ref{fig:f2}. On the CPU, the inference time increased on average by 7\% for InceptionV3 and 12\% for Unet. Although our analysis does not include OS as a decision variable it is interesting to note that there exists inference time variations between OS. This could be a component of the decision process alongside factors such as initial cost of the OS, availability of platform support, and running costs for OS (IT, administrative support cost, etc.).

\begin{figure}[t]
  \centering
  \subfloat[]{\includegraphics[trim=0cm 0cm 0cm 0cm, clip, width=2.5in]{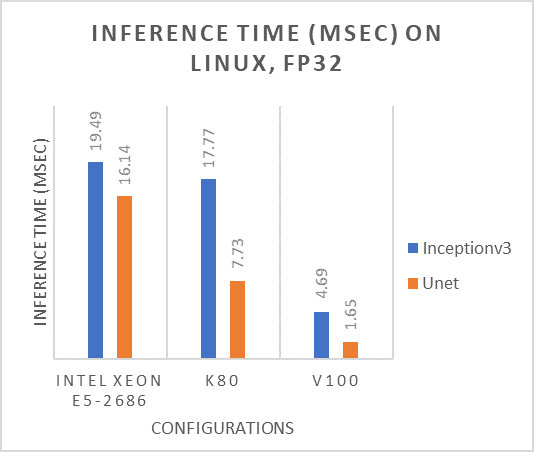}} 
  \subfloat[]{\includegraphics[trim=0cm 0cm 0cm 0cm, clip, width=2.5in]{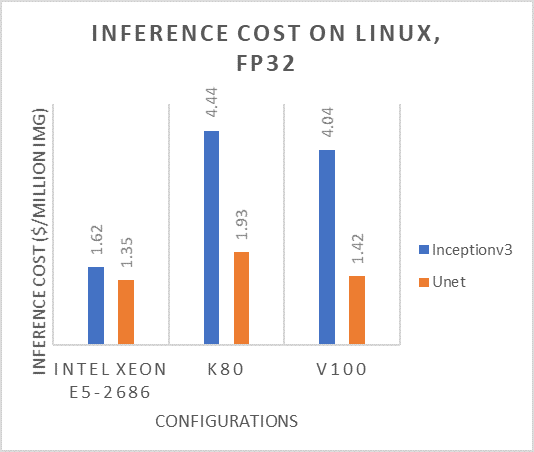}}   
  \caption{{Comparison of inference time (a) and inference cost (b) on Linux for single precision.}}
  \label{fig:f45}
\end{figure}

In our simulation, for segmentation with a Unet model, we would recommend using NVIDIA V100. But for InceptionV3, without any inference time constraints, a CPU instance works fine, and we would recommend using Intel Xeon for a lower inference cost. This interestingly shows that results vary for different models and data characteristics. Indeed, some tasks might depend on memory bandwidth, especially Recurrent Neural Networks that are out of scope here, whereas others would rely on computational speed. Hence, this analysis process proves itself relevant and useful to drive these situation-dependent decisions.

\begin{table}
  \caption{Inference time and cost comparison across all configurations}
  \label{inference-table1}
  \centering
  \begin{tabular}{llllll}
    \toprule
    \multicolumn{2}{c}{Results on Linux} &\shortstack{NVIDIA\\K80} &\multicolumn{2}{c}{\shortstack{NVIDIA\\V100}}  & Intel Xeon  \\
    \cmidrule(r){3-6}
    &  & FP32 &FP32&  FP16$*$&  FP32\\
    \midrule
    
    \multirow{2}{*}{InceptionV3 } & Inference time(msec/img) &17.77&  4.69& 3.34  &19.49\\
    \cmidrule(r){2-6}
    &Inference cost (\$/million images)& 4.44&  4.04  &2.88 &1.68\\
    \midrule
    \multirow{2}{*}{Unet} &Inference time(msec/img) &7.73 & 1.65& 1.34& 16.14\\
    \cmidrule(r){2-6}
    &Inference cost (\$/million images)& 1.93&  1.42& 1.15& 1.35\\
    \bottomrule
  \end{tabular}
\end{table}

\section{Discussion}

\subsection{Impact of speed limit on cost}
\begin{figure}[t]
  \centering
  \subfloat{\includegraphics[trim=0cm 0cm 0cm 0cm, clip, width=4.2in]{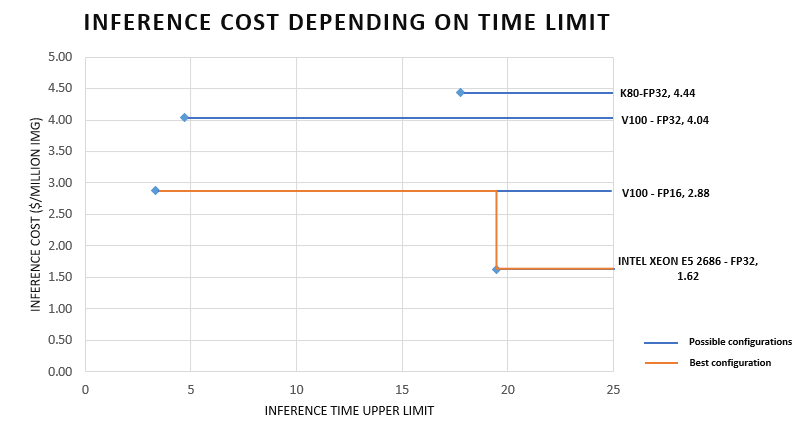}} 
  \caption{{Inference cost and hardware decision depending on time limit for InceptionV3}}
  \label{fig:Deicision}
\end{figure}

Our experiments allowed us to select the optimized configuration for deployment of each benchmarked model to reduce costs depending on the limit set for inference time. Figure \ref{fig:Deicision} shows the results for the InceptionV3 classifier and highlights three different situations: no optimal solution, fast solution for a higher price, and lower price for a longer inference time. In this example, the minimal inference time is 3.34 ms/image for V100 with FP16. If this speed was still not sufficient, the model would have to be modified to fit the deployment constraints. If the inference time limit is below 19.5 msec per image, the optimal hardware would be V100 with FP16 precision, if the 1\% drop in accuracy is acceptable, else V100 with FP32 precision, for a cost around \$2.88/million inferences, or \$4.04 for higher precision. K80 GPUs were never found to be optimal for any time constraints. If the deployment situation allows for an inference speed around 20 msec per image, the Intel XEON CPU is the optimal solution and reduces inference price by 35\%. Hence, depending on the constraints of the deployment situation, the optimal hardware changes, which impacts the cost for inferencing.

\subsection{Additional Factors}
Hardware decisions are inserted into the product development cycle. In healthcare, this cycle lasts many years and upgrades happen at regular intervals. This can be an opportunity to change configurations, but initial decisions are very impactful and must be made with all information at hand. The support and flexibility of the hardware provider is another major concern if the hardware is to be packaged into medical devices. Comparing documentation and understanding the user-base of the hardware is an informative indicator to consider when making such decisions.

Our decision model is based on very straight-forward criteria and relies on the axiom that decision-makers have no other constraints than inference time and costs. We do not consider cloud vs on-premise hardware, for instance. In healthcare, many additional factors would impact this choice. Even if legislation constraints are overcome, cloud-based hardware calls for specific systems to load and process data, and this set-up cannot be ignored. Changing the already implemented processes to make them cloud compatible is time-consuming and difficult. A comparison between cloud and on-premise devices might be valuable but would require considering other costs and organizational impacts. 

\subsection{Specific Configurations}
Some users in healthcare face constrained hardware configuration options, for example machines which contain only AMD GPUs. We benchmarked AMD Radeon E9260 GPU performances on Windows to determine the optimal path between CPU and GPU in these situations. We conducted the experiments with the same models as before. WinML was used to inference with the ONNX \cite{onnx} model format, since no other inference accelerators support AMD in Windows. 

We used the performances of a TensorFlow frozen model on the CPU as a baseline. Leveraging AMD via WinML allowed us to reduce inference time by 41\% for InceptionV3 and 75\% for Unet (See Table \ref{AWS-table}). Nevertheless, for CPU inferencing, WinML performs significantly worse in comparison to OpenVINO in terms of inference time. 

The selection of the optimal inference accelerator depends on the model and task at hand. In our experiments, we observe that optimizing the classification task with OpenVINO is the most optimal path, whereas for the segmentation task WinML should be used.

%%%%%%%%%%%%%%
\begin{table}
  \caption{Inference time comparison for different model optimizers on a local machine}
  \label{AWS-table}
  \centering
  \begin{tabular}{llllll}
    \toprule
    \multicolumn{2}{c}{Remote Window Machine} & TensorFlow  &OpenVINO&  \multicolumn{2}{c}{WinML}\\
    \cmidrule(r){3-6}
    & & \shortstack{Intel Core\\ i5-4590S CPU}&\shortstack{Intel Core\\ i5-4590S CPU}&\shortstack{Intel Core\\ i5-4590S CPU}  &\shortstack{AMD Radeon\\ E9260}\\
    \midrule
        \multirow{2}{*}{\shortstack{Inference time\\ (ms/img)}} &InceptionV3 &  92.71 &40.47  &256.56&  54.95\\
        \cmidrule(r){2-6}
        &Unet & 87.92 & 41.98 & 155.98  &22.34\\
    \bottomrule
  \end{tabular}
\end{table}

%%%%%%%%%%%%%%%%%%%%%%%%%%%%%

\section{Conclusion}
Our analysis highlights the importance of hardware decision on model deployment. This is a real challenge in healthcare where AI models are deployed on device. We observe that in some situations, constraints can become opportunities. Indeed, requirements in terms of inference speed can be leveraged to choose the optimal configuration and reduce costs. To this end, we present a decision process to determine what hardware is the most appropriate to a specific use-case, depending on model and data. Understanding the impact of the decision on inference speed and cost is the first step towards optimizing deployment. Nonetheless, some healthcare use cases such as live support and vital signals monitoring would still require the configuration with fastest inference. Because of its complexity, model deployment across devices remains an open-ended challenge, where optimal solutions vary depending on the task to solve. 

\section*{Acknowledgments}
We acknowledge Bangalore ATG team for providing us with the data and U-Net segmentation model for experiments. We thank Harald Deischinger and team for providing us with the hardware setup.

\bibliographystyle{unsrt}
\bibliography{benchmark}

\end{document}